\documentclass[aps,prl,numerical,linenumbers,superscriptaddress,showpacs,floatfix,reprint]{revtex4-1}
\usepackage[dvipdfm]{graphicx}
\usepackage{amsmath,amssymb,amsfonts}
\usepackage{color}

\newcommand{\be}{\begin{eqnarray}}
\newcommand{\ee}{\end{eqnarray}}

\def\v2{\mbox{$v_2$}}

\newcommand{\mean}[1]{\left\langle #1 \right\rangle}

\begin{document}


%
\title{Does quark number scaling breakdown in Pb+Pb collisions at $\sqrt{s_{NN}} = 2.76$~TeV?
}
%
%
%
\author{ Roy~A.~Lacey}
\email[E-mail: ]{Roy.Lacey@Stonybrook.edu}
\affiliation{Department of Chemistry, 
Stony Brook University, \\
Stony Brook, NY, 11794-3400, USA}
\author{Yi Gu} 
\affiliation{Department of Chemistry, 
Stony Brook University, \\
Stony Brook, NY, 11794-3400, USA}
\author{ X. Gong} 
\affiliation{Department of Chemistry, 
Stony Brook University, \\
Stony Brook, NY, 11794-3400, USA}
\author{D. Reynolds} 
\affiliation{Department of Chemistry, 
Stony Brook University, \\
Stony Brook, NY, 11794-3400, USA}
\author{ N.~N.~Ajitanand} 
\affiliation{Department of Chemistry, 
Stony Brook University, \\
Stony Brook, NY, 11794-3400, USA}
\author{ J.~M.~Alexander}
\affiliation{Department of Chemistry, 
Stony Brook University, \\
Stony Brook, NY, 11794-3400, USA}
\author{A. Mawi}
\affiliation{Department of Chemistry, 
Stony Brook University, \\
Stony Brook, NY, 11794-3400, USA}
\author{A.~Taranenko}
\affiliation{Department of Chemistry, 
Stony Brook University, \\
Stony Brook, NY, 11794-3400, USA} 
%



\date{\today}


\begin{abstract}

	The anisotropy coefficient $v_2$, for unidentified and identified 
charged hadrons [pions ($\pi$), kaons ($K$) and protons ($p$)] 
measured in Au+Au collisions at $\sqrt{s_{NN}}= 0.20$ TeV (RHIC) and Pb+Pb collisions 
at $\sqrt{s_{NN}}= 2.76$ TeV (LHC), are compared for several collision centralities
($\text{cent}$) and particle transverse momenta $p_T$. In contrast to the measurements
for charged hadrons, the comparisons indicate a sizable increase of $v_2(p_T)$ for $\pi,K$ and $p$, 
as well as a blueshift of proton $v_2(p_T)$, from RHIC to LHC. When this blueshift is accounted for, 
the LHC data [for $\pi$, $K$, $p$] show excellent scaling of $v_2({K\!E}_T)$ with 
the number of valence quarks ($n_q$), for a broad range of transverse 
kinetic energies (${K\!E}_T$) and collision centralities.
These observations suggest a larger mean sound speed $\left< c_s(T) \right>$ for 
the plasma created in LHC collisions, and significant radial flow generation 
after its hadronization. 
	
\end{abstract}


\maketitle


Anisotropic flow measurements for identified and unidentified charged hadrons are currently
being pursued at both the Relativistic Heavy Ion Collider and the Large Hadron Collider, to aid 
investigations of the temperature ($T$) dependence of the equation of state (EOS) 
and other transport properties of the hot and dense plasma  
produced in energetic heavy ion collisions 
\cite{Aamodt:2010pa,RaimondQM:2011yba,Lacey:2011av,Lacey:2011ug,Adare:2011tg,Adare:2012vq,
Phnxprelim:2012,Gong:2011zz,ATLAS:2012at,Chatrchyan:2012ta,STAR:2012ku}.
An important lever arm for these efforts is the measured energy density increase 
of more than a factor of three from RHIC to LHC \cite{Chatrchyan:2012mb}. 
This increase could result in a change in the mean specific shear 
viscosity $\left< \frac{\eta}{s}(T) \right>$
(the ratio of shear viscosity $\eta$ to entropy density ($s$)),
as well as a change in the value of the 
mean sound speed $\left< c_s(T) \right>$. Either could have a significant 
influence on the expansion dynamics, which in turn, influences the magnitude and trend 
of anisotropic flow. Thus, a crucial question is the extent to which flow measurements
for identified and unidentified charged hadrons differ from RHIC to LHC, and 
whether any characterizable difference reflects the sizable increase 
in energy density from RHIC to LHC?

Flow manifests as an anisotropic emission of particles in the 
plane transverse to the beam direction \cite{Ollitrault:1992bk,Lacey:2001va,*Snellings:2001nf},    
and is often characterized via Fourier decomposition of the measured azimuthal 
distribution for these particles;
\begin{equation}
\frac{dN}{d(\phi-\Psi_n)} \propto \left(
1 + \sum_{n=1} 2 \, v_n \, \cos(n[\phi-\Psi_n]) \right),  
\label{eq:1}
\end{equation}
where $\phi$ is the azimuthal angle of an emitted particle, 
$v_n = \mean{ \cos(n[\phi - \Psi_{n}])}, n=1,2,3,...$
and the $\Psi_n$ are the generalized participant event
planes at all orders for each event.
Characterization can also be made via the pair-wise distribution 
in the azimuthal angle difference ($\Delta\phi =\phi_1 - \phi_2$) between particle pairs 
with transverse momenta $p^a_{T}$ and $p^b_{T}$ 
(respectively) \cite{Lacey:2001va,Adcox:2002ms,ATLAS:2012at};
\begin{equation}
\frac{dN^{\text{pairs}}}{d\Delta\phi} \propto \left( 1 + \sum\limits_{n = 1} 
2v_n^a(p_T^a)v_n^b(p_T^b)\cos(n\Delta\phi) \right).
\label{eq:2}
\end{equation}
A sizable pseudorapidity-gap ($\Delta\eta'$), 
which serves to minimize  possible non-flow effects, 
is usually imposed between the particles in each pair to ensure consistency between 
the $v_n$ coefficients obtained via Eqs. \ref{eq:1} and \ref{eq:2} \cite{ATLAS:2012at,Chatrchyan:2012xq}.

Current RHIC $v_n$ measurements can be understood in terms of an eccentricity-driven 
hydrodynamic expansion of the high energy density quark gluon plasma (QGP) created 
in the overlap zone of the two colliding Au nuclei \cite{Heinz:2001xi,Teaney:2003kp,Huovinen:2001cy, 
Hirano:2002ds,Andrade:2005tx,Nonaka:1999et,Romatschke:2007mq,Song:2008hj,Niemi:2008ta,
Dusling:2007gi,Bozek:2009mz,Peschanski:2009tg,Denicol:2010tr,Holopainen:2010gz,Schenke:2010rr}.
That is, a finite eccentricity $\varepsilon_n$, drives uneven pressure gradients in- and out 
of the $\Psi_n$ event plane, and the resulting expansion of the plasma, 
modulated by a relatively small $\left< \frac{\eta}{s}(T) \right>$ value, 
leads to anisotropic particle emission about this plane. The observation that
$v_n({K\!E}_T)/(n_q)^{n/2}$ vs. ${K\!E}_T/(n_q)$ gives a universal curve for a broad 
spectrum of particle species [termed Quark Number Scaling (QNS)] 
\cite{Lacey:2006pn,Adare:2006ti,Lacey:2011av}, 
also gives a strong indication that anisotropic flow at RHIC develops primarily in the 
partonic phase, and is not strongly influenced by the subsequent hadronic phase.

The LHC $v_n$ measurements can also be understood in terms of an eccentricity-driven 
hydrodynamic expansion of the QGP created at a much 
higher energy density, in Pb+Pb collisions. However, recent comparisons 
of RHIC and LHC $v_2(p_T)$ data for unidentified charged hadrons, 
have indicated a striking similarity between the two 
sets of measurements \cite{Aamodt:2010pa,Lacey:2010ej}. 
This similarity posed an initial conundrum because the significant 
increase in energy density for LHC collisions, is expected to 
influence the expansion dynamics and hence, the magnitude of $v_{2}(p_T)$. 
Initial estimates of $\left< \frac{\eta}{s}(T) \right>$ 
from LHC charged hadron data, have not indicated a sizable change from 
RHIC to LHC  \cite{Lacey:2010ej,Schenke:2011tv,Qiu:2011hf}. However, in contrast to RHIC results, 
tests for QNS with LHC data for identified charged hadrons, have indicated an 
apparent breakdown of this scaling \cite{RaimondQM:2011yba}.	
	
%
\begin{figure*}[t]
\includegraphics[width=1.0\linewidth]{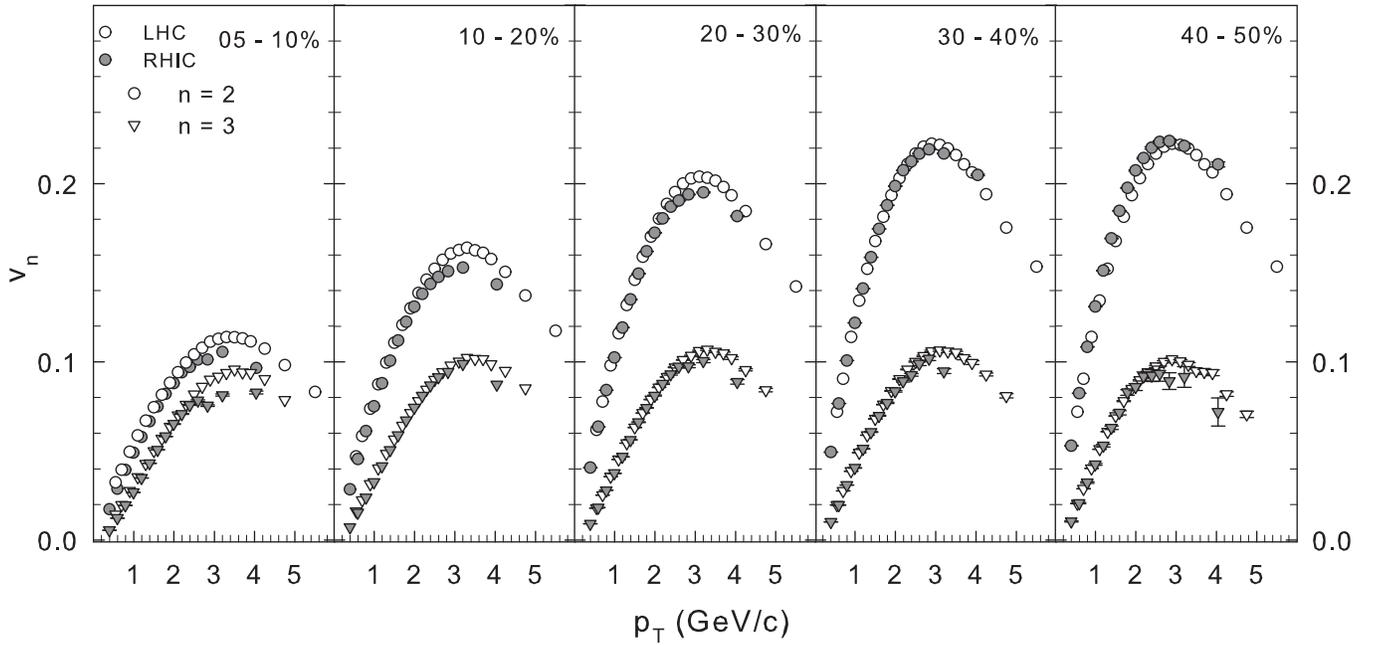}
%
\caption{ Comparison of $v_{2,3}(p_T)$ for charged hadrons obtained in 
Au+Au collisions at $\sqrt{s_{NN}}= 0.20$ TeV (RHIC) and Pb+Pb collisions 
at $\sqrt{s_{NN}}= 2.76$ TeV (LHC). The data are taken from from 
Refs. \cite{Adare:2011tg,Gong:2011zz} and \cite{ATLAS:2012at}.
}
\label{Fig1}
\end{figure*}
%
%

	In this work we present comparisons of RHIC and LHC flow measurements for both unidentified and 
identified charged hadrons, to further investigate whether the sizable increase 
in energy density from RHIC to LHC, signals a possible change in the expansion dynamics. 
We also study how such a change could manifest as a breakdown of quark number scaling. 
	
The double differential measurements $v_2(p_T,\text{cent})$ employed in our comparisons
are taken from the unidentified charged hadron results reported by 
the PHENIX  \cite{Adare:2011tg,Gong:2011zz} and ATLAS \cite{ATLAS:2012at} collaborations, 
as well as the measurements reported for identified charged hadrons by 
the PHENIX \cite{Adare:2012vq,*Phnxprelim:2012} and ALICE collaborations \cite{RaimondQM:2011yba}.

To initiate our comparisons, we show RHIC and LHC $v_{2,3}(p_T)$ measurements for unidentified 
charged hadrons ($h$) for several centrality selections in Fig. \ref{Fig1}. 
A comparison of the $v_2(p_T)$ measurements indicates good agreement between the magnitude and trends 
of both data sets for a broad range of $p_T$ and centralities, as 
previously reported \cite{Aamodt:2010pa,Lacey:2010ej}. The comparison also indicates 
that the observed similarity between RHIC and LHC charged hadron measurements 
extends to the higher harmonics.

\begin{figure}[!]
\includegraphics[width=1.0\linewidth]{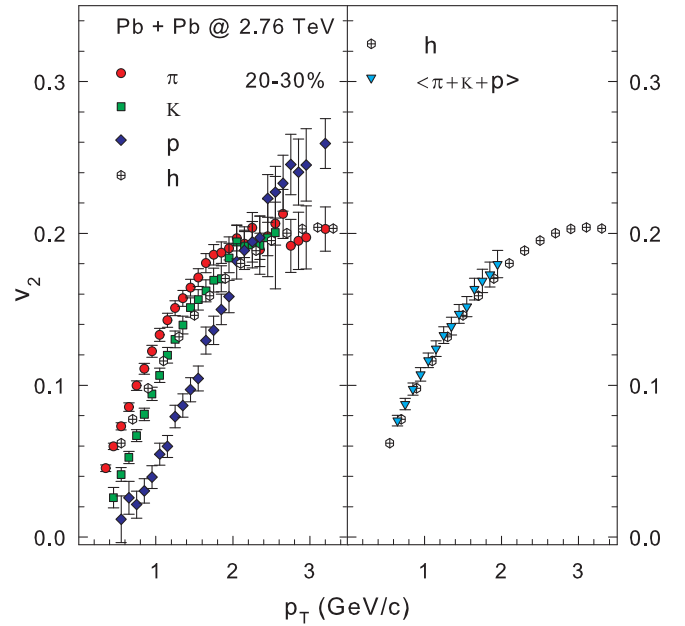}
\caption{(color on line) (a) Comparison of $v_2(p_T)$ vs. $p_T$ for $\pi,K,p$ and 
unidentified charged hadrons $h$. 
(b) Comparison of $v_2(p_T)$ for $h$ and the weighted average of the 
values for $\pi,K$ and $p$. The data for identified and unidentified charged hadrons, 
are from the ALICE \cite{RaimondQM:2011yba} and ATLAS \cite{ATLAS:2012at} collaboration 
respectively. Results are shown for the 20-30\% most 
central Pb+Pb collisions.
}
\label{Fig2}
\end{figure}

The $v_2(p_T)$ results for charged hadrons are actually a weighted average of the values 
for identified charged hadrons. Consequently, one can test for consistency between 
the measured values of $v_2(p_T)$ for identified and unidentified charged hadrons.
Such a consistency check is shown for LHC data in Fig. \ref{Fig2}. 
The left panel of Fig. \ref{Fig2} shows that, while $v_2(p_T)$ for $K$ and $h$ 
are similar for most of the $p_T$ range, they are 
significantly different from the the values for $\pi$ and $p$. 
The right panel of Fig. \ref{Fig2} shows that an appropriate averaging of the same  
$v_2(p_T)$ values for $\pi,K$ and $p$ (with weights derived from the measured  
$p/\pi$ and $K/\pi$ ratios), gives average values which are essentially the same as 
those for $h$.  

%

\begin{figure}[t]
\includegraphics[width=1.0\linewidth]{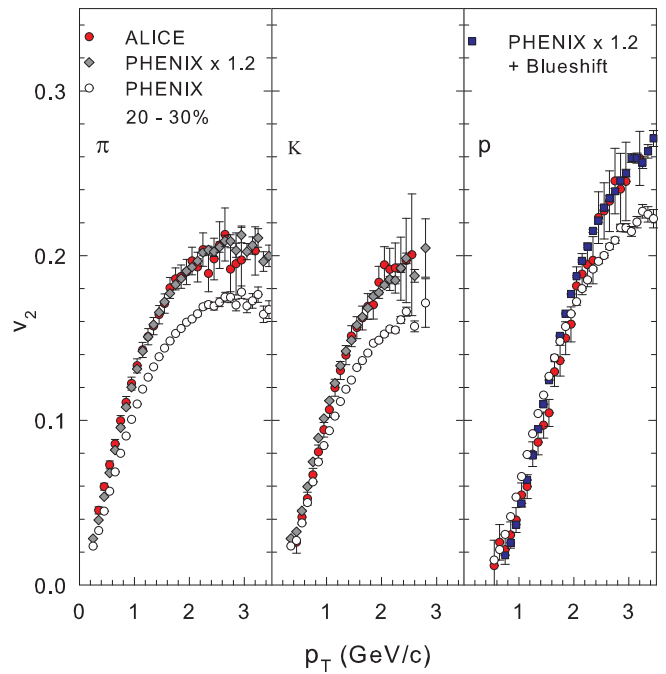}
\caption{(color on line) Comparison of PHENIX and ALICE data for $v_2(p_T)$ vs. $p_T$ 
for $\pi,K$ and $p$ as indicated.
Results are shown for the 20-30\% most central collisions.
}
\label{Fig3}
\end{figure}

	Given the substantial differences between the LHC $v_2(p_T)$ values for $\pi,K$ and $p$
shown in the left panel of Fig. \ref{Fig2}, it is important to ask whether 
the agreement observed between RHIC and LHC data for $h$ (Fig. \ref{Fig1}) translates 
to a similar agreement between RHIC and LHC measurements for $\pi,K$ and $p$ (respectively)?
Fig. \ref{Fig3} compares the RHIC (open circles) and LHC (filled red circles) $v_2(p_T)$ 
values for $\pi,K$ and $p$ for the 20-30\% most central events. 
The values for $\pi$ and $K$ give a clear indication that 
the LHC values are approximately 20\% larger than the RHIC values. This is 
confirmed by the excellent agreement between RHIC and LHC measurements after the 
scale factor $\sim 1.2$ is applied to the RHIC data 
(filled circles in left and middle panels of Fig. \ref{Fig3}). 

	For $p_T \agt 2.5$~GeV/c, the results for protons, shown in the right panel 
of Fig. \ref{Fig3}, also hint at a 20\% difference between the RHIC and LHC values. 
For lower $p_T$ ($p_T \alt 2.0$~GeV/c) however, the RHIC $v_2(p_T)$ values appear to be 
larger than the LHC values. We attribute this inversion to a 
small blueshift of the LHC values. Such a blueshift has been 
observed in recent viscous hydrodynamical calculations for LHC collisions 
\cite{Heinz:2011kt}, and can be linked to a sizable increase in the magnitude of 
the radial flow generated in these collisions, especially in the hadronic phase. 
Here, the blueshift is confirmed by the excellent agreement obtained 
between the proton measurements, when the RHIC data are scaled by the 
factor $\sim 1.2$ (as for $\pi$ and $K$) and then blueshifted by $\sim 0.2$~GeV/c
(filled squares in right panel of Fig. \ref{Fig3}).
Similarly good agreement between RHIC and LHC measurements  
were obtained for other centrality selections, with essentially the same 
blueshift value. However, a larger (smaller) scale factor was required for more 
central (peripheral) collisions, as might be expected from the change in 
energy density with collision centrality.
%
%

	The results from the comparisons shown in Fig. \ref{Fig3}, suggest that 
the agreement observed between the charged hadron measurements in Fig. \ref{Fig1},
may be inadvertent. Thus, the comparisons for charged hadrons might not convey all of 
the essential information about the expansion dynamics. By contrast, 
the observed increase in $v_2(p_T)$ from RHIC to LHC for identified 
charged hadrons (Fig. \ref{Fig3}), suggests that the expansion 
dynamics in LHC collisions is driven by a larger 
mean sound speed $\left< c_s(T) \right>$ for the plasma created in these collisions. 
Such an increase in $\left< c_s(T) \right>$ could result from the sizable increase 
in energy density from RHIC to LHC.    

\begin{figure*}
\includegraphics[width=1.0\linewidth]{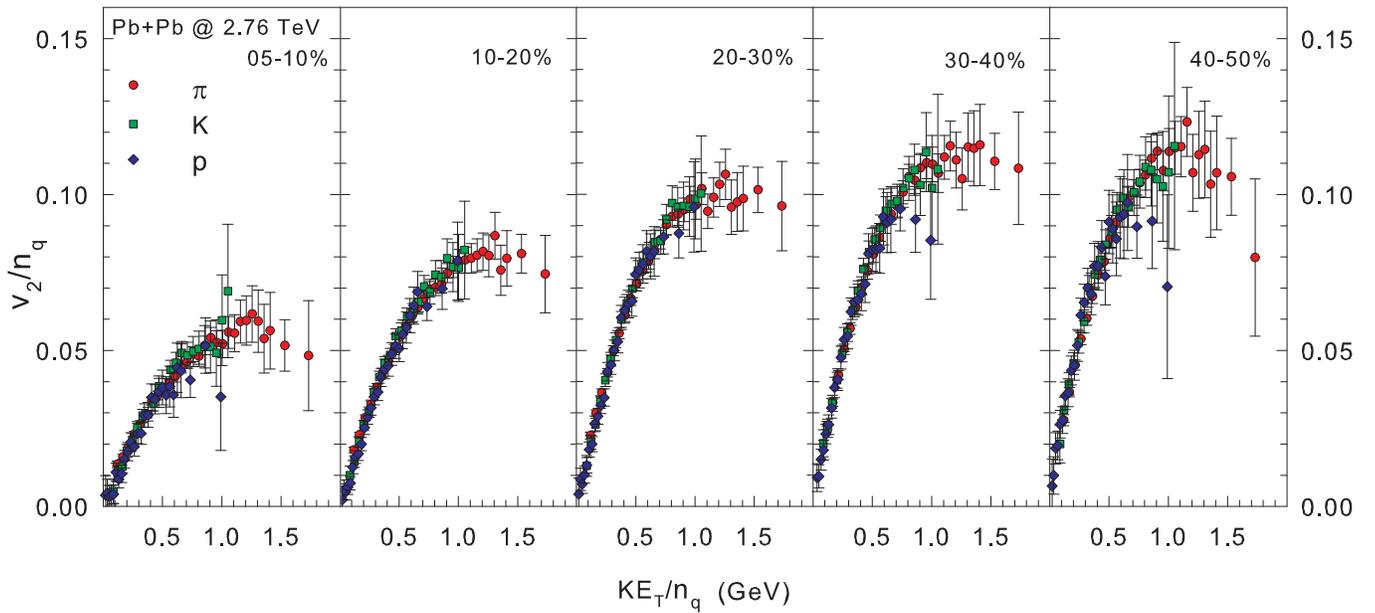} 
\caption{(Color on line) ${v_2(K\!E_T)}/{n_q}$ vs. $K\!E_T/n_q$ for pions, kaons and protons,  
after correcting for the proton blueshift (see text). Results are show for several 
centrality selections as indicated.
}
\label{Fig4}
\end{figure*}

	The blueshift inferred for proton $v_2(p_T)$ in LHC collisions is incompatible
with quark number scaling. Thus, it provides a straightforward 
explanation for the observed failure of this scaling, when applied to LHC data for 
identified charged hadrons ($\pi, K$ and $p$) \cite{RaimondQM:2011yba}. 
An appropriate correction for this blueshift would of course, lead 
to a restoration of quark number scaling. This is  
demonstrated for LHC data in Fig. \ref{Fig4}, for a broad range of 
centrality selections. For these plots, the $v_2(p_T)$ data for 
protons were redshifted by $\sim 0.2$~GeV/c for each centrality selection 
[prior to QNS scaling] to account for the blueshift (cf. right panel of Fig. \ref{Fig3}) with 
the same magnitude. Fig. \ref{Fig4} shows that this procedure leads to 
excellent quark number scaling of the LHC data for identified charged hadrons, 
and confirms that partonic flow still dominates for LHC collisions.
Note as well, that the magnitudes of the quark number scaled values 
of ${v_2(K\!E_T)}/{n_q}$ are significantly larger than those observed 
at RHIC.

%
%
%


In summary, we have performed detailed comparisons of RHIC and 
LHC flow data for unidentified and identified charged hadrons.
In contrast to the agreement observed between the RHIC and LHC data
sets for unidentified charged hadrons, $v_2(p_T)$ for $\pi,K$ and $p$ 
indicate a sizable increase from RHIC to LHC. 
This increase is compatible with the larger mean sound 
speed $\left< c_s(T) \right>$, expected for the plasma created at a much 
higher energy density in LHC collisions. 
The comparisons also indicate a blueshift of LHC proton $v_2(p_T)$
relative to RHIC proton $v_2(p_T)$, possibly because of a sizable growth 
of radial flow in the hadronic phase for LHC collisions. 
When this blueshift is accounted for, excellent scaling 
of $v_2({K\!E}_T)$ with the number of valence quarks is observed [for $\pi$, $K$, $p$] 
for a broad range of transverse kinetic energies and collision centralities.
These results highlight the indispensable role of the measurements for identified 
particle species at both RHIC and the LHC, for studies of the temperature ($T$) 
dependence of the equation of state (EOS) and other transport properties. 

{\bf Acknowledgments}
We thank R. Snellings for providing the ALICE experimental data.
This research is supported by the US DOE under contract DE-FG02-87ER40331.A008. 
 


%
\bibliography{ncq_scaling_LHC} 
\end{document}